\documentclass[prl,aps,twocolumn,showpacs,superscriptaddress]{revtex4}
\usepackage{graphicx}
\usepackage{amsmath,amssymb,bm}

\newcommand{\beq}{\begin{equation}}
\newcommand{\eeq}{\end{equation}}
\newcommand{\barr}{\begin{eqnarray}}
\newcommand{\earr}{\end{eqnarray}}

\begin{document}
\title{Multipartite entanglement generation and
fidelity decay in disordered qubit systems}

\author{Simone Montangero}
\thanks{Corresponding author. E-mail: {\tt monta@sns.it}}
\affiliation{NEST-CNR-INFM $\&$ Scuola Normale Superiore, Piazza
             dei Cavalieri 7, 56126 Pisa, Italy}

\author{Lorenza Viola}
\affiliation{\mbox{Department of Physics and Astronomy, Dartmouth College, 
             6127 Wilder Laboratory, Hanover, NH 03755, USA}}

\begin{abstract}
We investigate multipartite entanglement dynamics in disordered
spin-1/2 lattice models exhibiting a transition from integrability to
quantum chaos.  Borrowing from the generalized entanglement framework,
we construct measures for correlations relative to arbitrary local and
bi-local spin observables, and show how they naturally signal the
crossover between distinct dynamical regimes.  Analytical estimates
are obtained in the short- and long-time limits.  Our results are in
qualitative agreement with predictions from random matrix theory and
relevant to both condensed-matter physics and to the stability of
quantum information in disordered quantum computing hardware.
\end{abstract}

\date{\today}

\maketitle 

Developing a quantitative understanding of the structural and
dynamical properties of entanglement in many-body quantum systems is a
critical challenge for both condensed-matter theory and quantum
information science. Low-dimensional disordered spin models offer, in
this context, an ideal testbed for theoretical analysis.  On one hand,
these systems are simple enough for analytic benchmarks to exist in
limiting situations, yet capable to demonstrate a broad typology of
complex quantum phenomena. The latter range from field- or
disorder-driven structural ground-state changes responsible for
quantum phase transitions~\cite{sachdev}, to dynamical crossovers from
integrable to non-integrable regimes and the emergence of quantum
chaos~\cite{haake}.  On the other hand, arrays of interacting
spin-1/2s naturally schematize quantum computing hardware as diverse
as semiconductor quantum dots~\cite{loss}, superconducting Josephson
qubits~\cite{schoen}, electron floating on He \cite{dykman}, and
optical lattices~\cite{pachos} -- disorder resulting from the
unavoidable presence of imperfections in the single-qubit energy
spacings and/or inter-qubit couplings.

Following~\cite{dima}, intense effort has been devoted to both assess
the impact of disorder on quantum computing
performance~\cite{flambaum,prl01,berman}, and to characterize
entanglement across the ensuing transition to quantum
chaos~\cite{santos,casati04}.  These studies point to the key role of
spectral properties, as captured by the {\em Local Density of States}
(LDOS)~\cite{note}, in determining the system stability against the
disorder.  Changes in the LDOS profile are ultimately responsible for
the existence of different dynamical regimes, as reflected by
qualitative changes in the rate of fidelity decay~\cite{fidelity}.
Numerical experiments confirm that the same regimes translate into
distinctive static~\cite{casati04} and dynamic ~\cite{prl03}
properties of pairwise entanglement, as quantified by
concurrence~\cite{wootters}.  While providing suggestive evidence,
such analyses are not fully satisfactory for two reasons.  First,
concurrence lacks a direct physical interpretation, hindering the
possibility to relate entanglement to spectral properties.  Second,
concurrence is a bipartite measure, preventing the quantification of
multipartite correlations which dominate in strongly coupled
scenarios.

In this Letter, we overcome the above limitations by exploiting {\em
Generalized Entanglement} (GE)~\cite{barnum} as a setting for defining
entanglement relative to arbitrary sets of observables.
In~\cite{somma}, GE measures constructed from algebras of fermionic
operators have been applied to the study of broken-symmetry quantum
phase transitions in exactly solvable models, notably the spin-1/2 XY
chain in a transverse field.  Here, we show how GE contributes to the
understanding of standard multipartite correlations between
distinguishable systems, by selecting {\em local} and {\em bi-local}
algebras of observables corresponding to single and pairs of qubits,
respectively.  Beside providing a transparent interpretation of the
results based on concurrence~\cite{prl03,casati04}, our approach
allows for quantitative insight about entanglement dynamics starting
from arbitrarily correlated initial states.  We quantify the influence
of LDOS properties in two limits: at short-time, by establishing a
direct link with fidelity decay; at long times, by relating the
entanglement saturation value to the participation ratio of the
asymptotic many-body state.

{\em The model}.-- We focus on a two-dimensional lattice of $n$
disordered spin-$1/2$ particles ($n$ even) described, in units
$\hbar=1$, by the following Hamiltonian:
$$H = \sum_{j=1}^n [\Delta + \delta_j] \sigma_z^{(j)} + \sum_{\langle
i,j \rangle} J_{ij} \sigma_x^{(i)}\sigma_x^{(j)}\equiv H_\Delta +
H_\delta +H_J\,, $$ where $\sigma_{\alpha}^{(i)}$, $\alpha \in \{
0,x,y,z\}$, $\sigma^{(i)}_0 =\mathbb{I}$, are Pauli operators, and
$\langle i,j \rangle$ stands for nearest-neighbor sites. Open boundary
conditions and energy units $\Delta \equiv 1$ are used. The parameters
$\delta_j, J_{ij}$ characterize the disorder in the on-site energy
splittings and two-body couplings, respectively.  We assume that
$\delta_j, J_{ij}$ are uniformly random over $[-\delta, \delta]$,
$[-J, J]$, with $\delta, J >0$.  The above model, which belongs to the
random transverse-field Ising lattice class~\cite{pastur}, was
recently used to describe a quantum register with static
imperfections~\cite{dima}. Understanding the influence of the latter
may be especially important for controlling entanglement in quantum
computers with always-on interactions~\cite{always} and one-way
quantum architectures~\cite{briegel}.

The spectrum of $H_\Delta$ is composed of $n +1$ levels, with energy
$E_B= \Delta (2 k - n)$ and degeneracy $N_B(k) = n!/[k! (n-k)!]$, $k=
0,\ldots, n$.  When $\delta, J \ll \Delta$, the degeneration is
removed and $n +1$ bands appears. The width $\Delta_B$ of the bands
depends on the ratio $J/\delta$~\cite{dima}.
In the limit where $J \ll \delta$, $\Delta_B \sim \delta \sqrt{n}$ as
the spread is led by the diagonal term $H_\delta$.  Correspondingly,
bands are Gaussian.  Inside each band, the effective number of states
coupled by $H_J$ is given by the width of the LDOS.  For $J \lesssim
\delta/n\equiv J_c$, $H_J$ couples few states, and the system is
slightly perturbed, while for $J > J_c$ quantum chaos sets in and the
unperturbed levels are coupled to a quasi-continuum set of states. The
LDOS becomes Lorentzian, with a width determined by the Fermi Golden
Rule (FGR), $\Gamma_F
= J^2 n /\delta$~\cite{dima}.  Increasing $J$, when $\Gamma_F \sim
\Delta_B$, that is for $J \sim J_E = \delta /n^{{1}/{4}}$, all the
levels inside a given band are mixed and the LDOS approaches the level
density. Thus, the LDOS becomes a Gaussian with width $\Gamma_E \sim
J$~\cite{dima}. In summary, three distinct regimes exist --
perturbative ($J < J_c$), FGR ($J < J_E$), and ergodic ($J_E < J <
\Delta$).  It has been shown that starting from an eigenstate
$|\psi_0\rangle$ of $H_\Delta$, the survival probability (fidelity
henceforth, as for initial eigenstates the two quantities coincide)
$F(t)=|\langle\psi_0| \psi_t\rangle|^2$ also follows three different
behaviors: ${F}(t)$ oscillates near one in the perturbative regime,
while it decays as an exponential or a Gaussian in the FGR and ergodic
regime, respectively~\cite{dima}.  The connection between LDOS shape
and fidelity decay has been unveiled in~\cite{flambaum} by means of a
perturbative relation between ${F}(t)$ and the Fourier transform of
the LDOS.

\begin{figure}[t]
    \centerline{\includegraphics[width=3.2in]{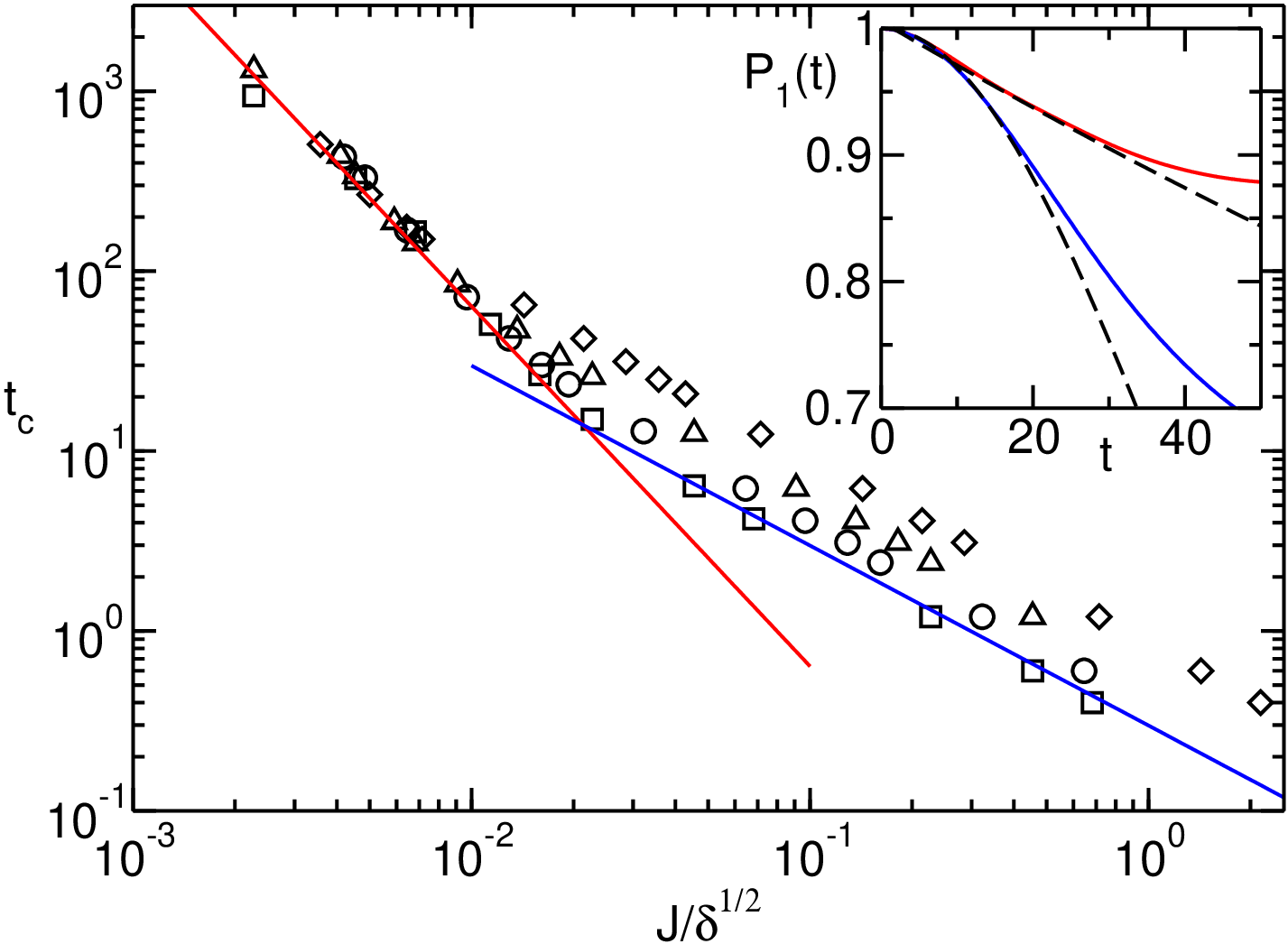}}
    \caption{(Color Online) Local purity critical time $t_c$ (${P}_1(t_c)=0.9$) 
     as a function of $J/\delta$ for $\delta=0.02,0.05,0.1,0.2$
     (diamonds, triangles, circles, squares).  To 
     smooth statistical fluctuations, here and in the following 
     we consider averages $\langle P_1(t)\rangle_D$ over a number 
     $N_r=10$ of disorder realizations. We assume $n=10$ 
     unless otherwise specified.  Inset: Time
     evolution of $P_1(t)$ in the FGR $J=\delta/10 = 0.01$
     (red) and ergodic $J=\delta=0.1$ (blue) regime starting from the
     state of Eq.~(\ref{istate}) for $n=14,n_B=0$. Dotted lines: 
     Exponential and Gaussian fits, as in Eq.~(\ref{regimes}).  }
    \label{fig1}
\end{figure}

{\em Local and bi-local purities}.-- An initial pure state
$|\psi_0\rangle$ evolving under $H$ remains pure for any fixed
disorder realization.  Accordingly, we consider pure-state
entanglement throughout.  The basic intuition underlying the GE
notion~\cite{barnum} is to quantify how entangled a state
$|\psi\rangle$ is {\em relative to an observable set ${\cal O}$} in
terms of how pure $|\psi\rangle$ remains upon restricting operational
access to ${\cal O}$.  While no single set ${\cal O}$ can exhaust the
complexity of multipartite entanglement, two simple choices will
illustrate the usefulness and flexibility of the algebraic approach.
First, we consider the set ${\cal O}={\cal O}_1$ generated by
arbitrary {\em local} observables that is, we probe the average
entanglement of each qubit with the rest of the lattice.  The
corresponding (time-evolved) {\em local purity} measure is defined as
\beq {P}_1 (|\psi_t\rangle) = \frac{1}{n} \sum_{\alpha=x,y,z}^{i=1,n}
|\langle\psi_t | \sigma_\alpha^{(i)} |\psi_t \rangle|^2\:.
\label{local}
\eeq 
For this choice of observables, GE coincides with {\em global
multipartite entanglement} as quantified by the Meyer-Wallach metric
$Q$~\cite{meyer,somma}, with $Q=1-{P}_1$~\cite{somma,brennen}.  As a
second observable set, whose physical motivation will become clear
later, we choose ${\cal O}={\cal O}_2$ generated by all the
observables acting on pairs of nearest neighbor spins.  We compute a
{\em bi-local purity} measure as \beq {P}_2 (|\psi_t\rangle) =
\frac{2}{3 \, n} \widetilde \sum_{\alpha,\beta=x,y,z,0}^{\langle i,j
\rangle} |\langle \psi_t | \sigma_\alpha^{(i)} \sigma_\beta^{(j)}
|\psi_t \rangle|^2\:,
\label{bilocal} 
\eeq where the tilde means that the identity term with
$\alpha=\beta=0$ is omitted.  ${P}_2$ may be thought as a
``coarse-grained'' version of ${P}_1$, resulting from ignoring the
fine structure given by arbitrary correlations within each pair. Both
measures are normalized to give one (zero) on states which are fully
separable (contain maximal GE) relative to the corresponding algebra.
Remarkably, both ${P}_1$ and ${P}_2$ are {\em directly measurable}
quantities in principle~\cite{brennen}.

{\it Initial separable state.}-- We first study multipartite
entanglement generation starting from a separable state in the central
band ($k=n/2$) that is, $|\psi_0\rangle =|010101 \dots 01\rangle
\equiv | c \rangle$, where $\{0,1\}$ label the states of each spin in
the computational basis and $c$ is the integer given by the
corresponding binary string.  The evolution of $P_1(t)$ under
different values of $J$ at fixed $\delta$ are depicted in the inset of
Fig.~\ref{fig1}: The decay of $P_1(t)$ clearly follows two different
dynamical laws.  Specifically, data in the FGR and in the ergodic
regime are described, respectively, by \beq
{P}_1^F(t) \approx e^{- C \, \Gamma_F t} \:, \hspace{5mm}
{P}_1^E(t) \approx e^{- C' \,\Gamma_E^2 t^2} \:, 
\label{regimes} 
\eeq where the constants $C,C'$ depend on the initial state and the
lattice coordination number (see e.g.~\cite{facchi}).  Let $t_c$ be
the time it takes for $P_{1}$ (or $P_2$) to reach the value $K$ e.g.,
${P}_{1}(t_c)= K$: We find that $t_F^c \sim 1/J^2$ in the FGR regime,
whereas $t_E^c \sim 1/J$.  These behaviors have been verified over a
wide range of disorder parameters, see Fig.~\ref{fig1}.

\begin{figure}[t]
   \centerline{\includegraphics[width=3.2in]{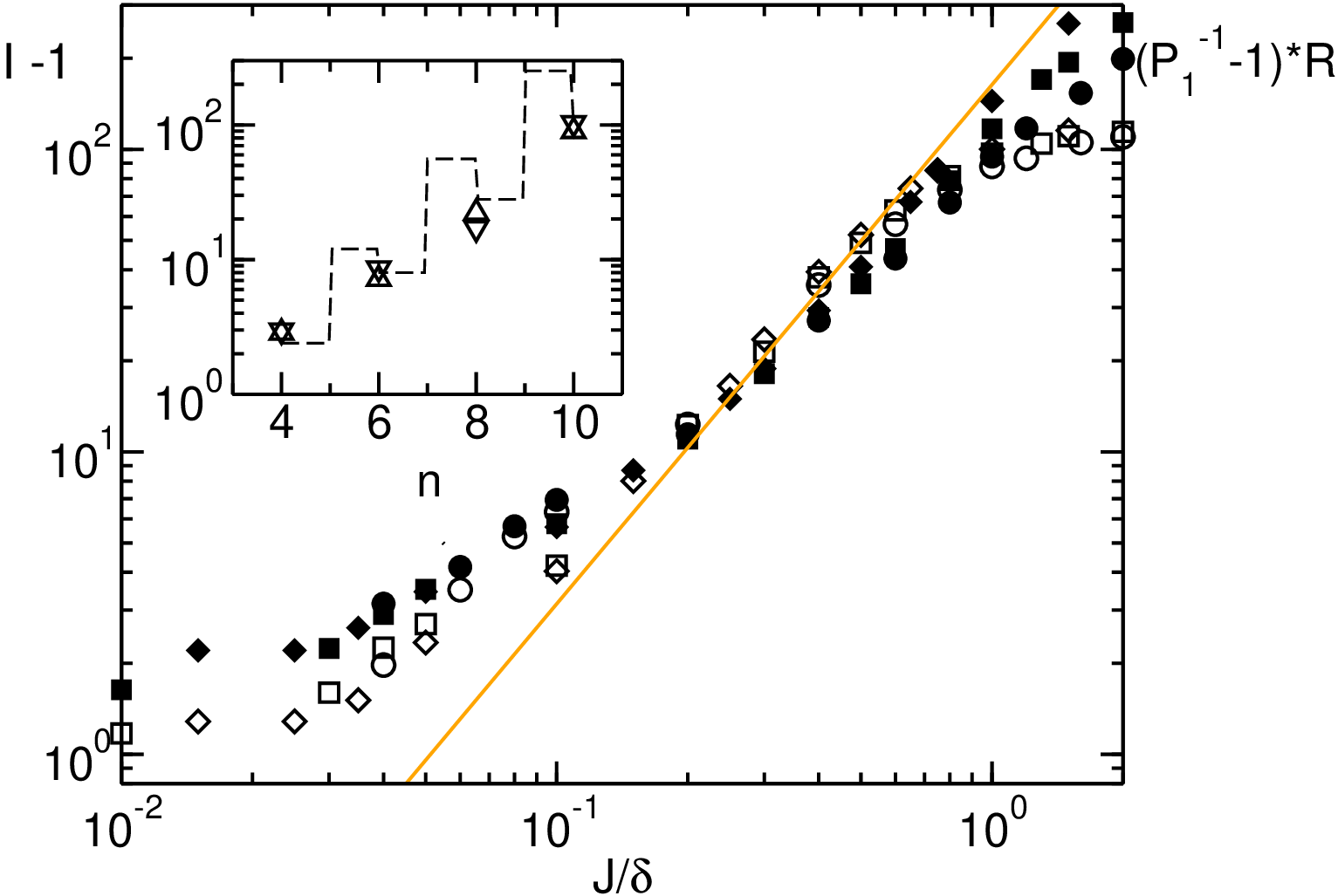}}
    \caption{(Color Online) Saturation values of I-1 (empty symbols) and of
    $({P}_1^{-1}-1)\cdot R$ (filled symbols) vs $J/\delta$ for
    $\delta=0.05$ (circles), $\delta=0.1$ (squares), $\delta=0.2$
    (diamonds), and rescaling factor $R=7$.  The straight line is
    proportional to $C (J/\delta)^2 N_B(k=n/2)$, $C \sim 0.7$.  Inset:
    Saturation value of the IPR (triangles up) and $P_1^{-1}$
    (triangles down) in the ergodic regime ($\delta=0, J=0.1$) as a
    function of $n$.  The dashed line gives \vspace*{-2mm}$N_B(n/2)$ vs $n$.
}
    \label{fig2}
\end{figure}

A simple physical interpretation of the above results follows from the
possibility to directly relate GE dynamics to fidelity decay for any
computational eigenstate $|c\rangle$.  While a full derivation will be
presented elsewhere, the key steps are (i) to realize that only
$\sigma_z^{(i)}$ observables contribute to the evolution of $P_1(t)$
(ii) to isolate the fidelity term $|\langle c|e^{-iHt}|c\rangle|^2$ in
the resulting $z$-purity.  This yields
$${P}_1(t) =
F(t)^2 + \frac{1}{n} \sum_{j=1}^{n}\left(
2 F(t) (-)^{\pi_j(c)} \alpha_j(t) + \alpha_j(t)^2 \right) \:,
$$
where $\alpha_j(t)= \sum_{p\ne c} |\langle p |e^{-iHt} |c\rangle |^2
(-)^{\pi_j(p)}$ and $(-)^{\pi_j(q)}$ $ =(-)^{\lfloor
  q/2^{j-1}\rfloor}$ is the parity of the $j$th qubit in the
computational state $|q\rangle$.  One can show that each term
$\alpha_j(t)$ is of order $\mathcal{O}((J/\delta)^2 t^2)$.  Thus,
according to the above equation, the connection between local purity
and fidelity decay (hence LDOS via the relation found
in~\cite{flambaum}) becomes exact in the limit $t\rightarrow 0$.  For
sufficiently short times, the first term still dominates and the
dynamics is governed by $F(t)^2$, whereby the two regimes of
Eq. (\ref{regimes}) arise.

\begin{figure}[t]
   \centerline{\includegraphics[width=3in]{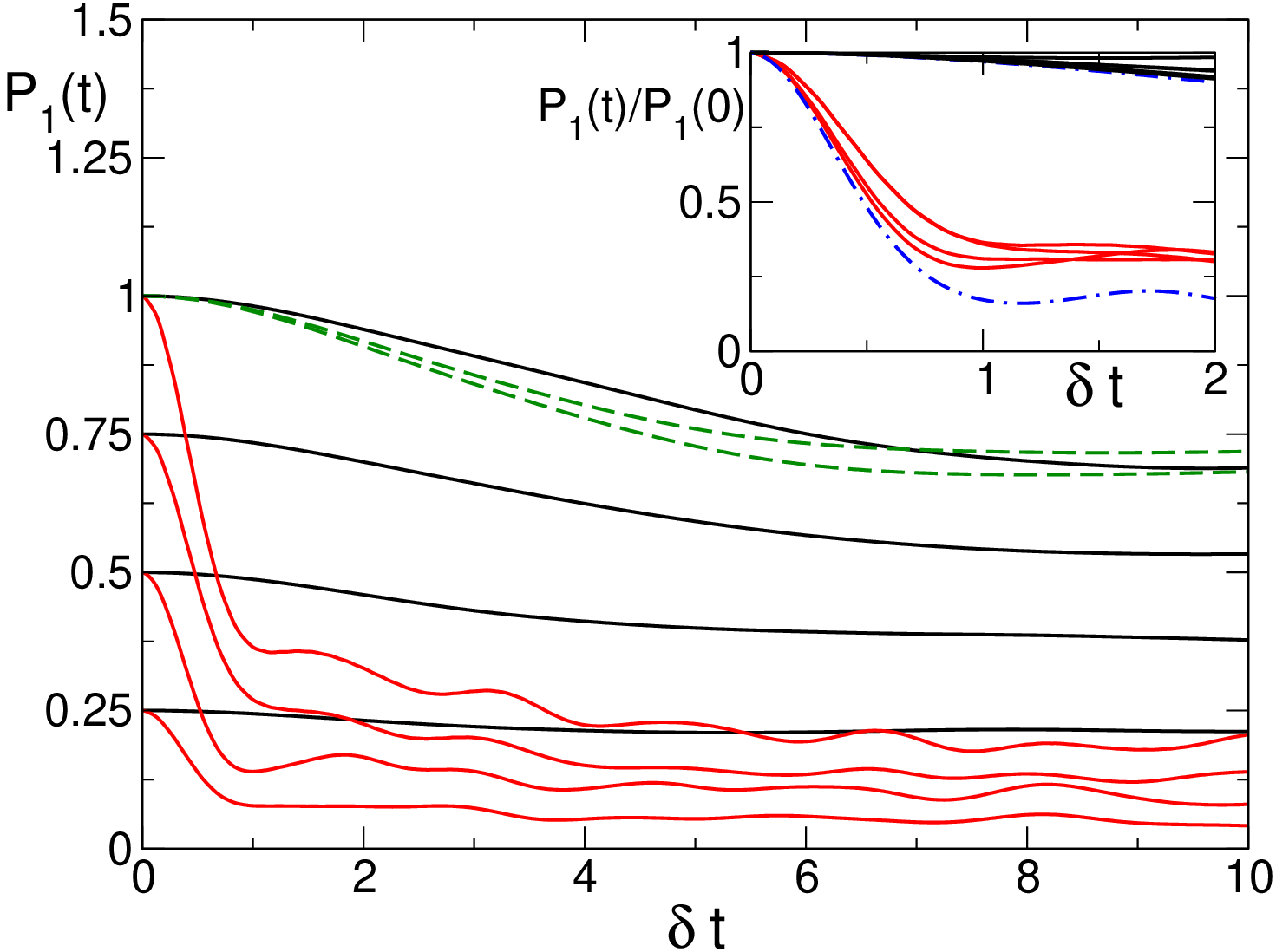}}
   \caption{(Color Online)
    Local purity ${P}_1(t)$ versus time in the FGR 
    (black lines) and ergodic regime (red lines) for different 
    initial states with $n_B=0,1,2,3$, $\delta=0.1$, and 
    $J=0.01$ (black curves), $J=0.1$ (red curves). Dashed lines:  
    Different separable initial states $|\psi_0\rangle = |0101010110\rangle,
    |010101010\rangle \otimes (|0\rangle + |1\rangle)/\sqrt{2}$ for $\delta=0.1, J=0.01$.
    Inset: The sames curves rescaled by 
    ${P}_1(0)$. Also shown (dot-dashed lines) is the local purity 
    decay in the ergodic and FGR regimes for an initial 
    $|{\tt W}\rangle$ state. 
    }
    \label{fig3}
\end{figure}

For times much longer than $\Gamma_{E,F}$, the asymptotic GE amount
may be estimated using random matrix theory~\cite{haake}. By working
within the Gaussian orthogonal ensemble, we represent the many-body
state $|\psi_\infty\rangle$ as a random superposition of $N_\infty$
unperturbed states within the central band that is, $|\psi_\infty
\rangle = \sum_{p=1}^{N_\infty} w_p |p\rangle$, with amplitudes $w_p$
satisfying the normalization constraint. For sufficiently large
$N_\infty$, correlations between different components may be
neglected, and the probability density of each component is well
approximated by the Porter-Thomas distribution, $P_{\tt
PT}({|w_p|^2})=(2\pi N_\infty |w_p|^2)^{-1/2} \exp(-N_\infty
|w_p|^2/2)$.  Evaluating the disorder-averaged asymptotic local purity
yields $\langle P_1^\infty \rangle_D = \langle P_1^\infty \rangle_{\tt
PT}= 2/N_\infty $, consistent with the scaling ${3}/({N_\infty +1})$
predicted for a {\em random} pure state under the invariant Haar
measure~\cite{scott}.

Let the degree of delocalization of $|\psi_\infty\rangle$ be
quantified by the {\em Inverse Participation Ratio}
(IPR)~\cite{haake,dima,casati04} that is, I= $1/\xi$, where the
asymptotic participartion number $\xi^\infty= \sum_p |w_p|^4$.  By
estimating $\left\langle \xi^\infty \right\rangle_D = 3/N_\infty$,
$\langle P_1^\infty \rangle_D = 2/3 \langle \xi^\infty \rangle_D$
-- thus, {\em GE directly quantifies delocalization in this regime}.
Similarly, $\langle \mbox{I}^\infty \rangle_D \approx N_\infty/3$,
up to deviations of the order ${\cal O}(N_\infty^{-1/2})$, consistent
with Gaussian fluctuation statistics~\cite{haake}.  Because the
dynamically accessible states are determined by the ratio between the
LDOS width and the bandwidth, we further estimate $N_\infty \approx
(\Gamma_F/\Delta_B) N_B$ for any initial eigenstate in the central
band and in the FGR regime, whereas the IPR is constant in the ergodic
regime.  These predictions have been confirmed numerically (see
Fig. \ref{fig2})~\cite{ynote}.  Thus, \beq \langle P_1^\infty
\rangle_D \propto \frac{1}{\langle \mbox{I}^\infty\rangle_D} \propto
\frac{\Delta_B}{ \Gamma_F N_B} \:.
\label{satp}  
\eeq As shown in Fig.~\ref{fig2}, Eq. (\ref{satp}) qualitatively
agrees with data over an extensive range of parameters.  Note that
these asymptotic results, as well as the initial decay laws of
Eq. (\ref{regimes}), are valid for the dynamics of {\em any} initial
separable state, not just of a computational state (see
Fig.~\ref{fig3}).

\begin{figure} 
    \centerline{\includegraphics[width=2.8in]{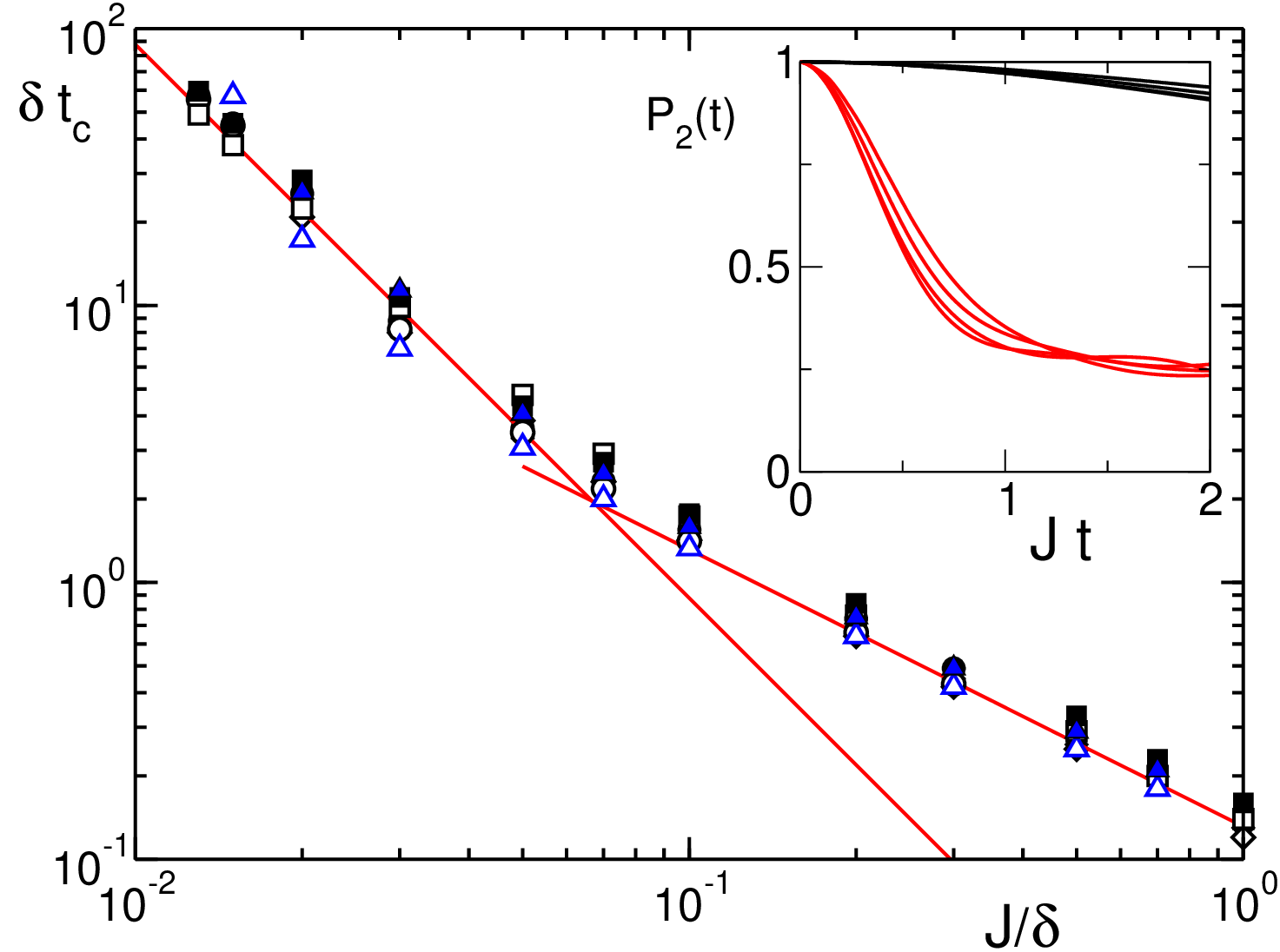}}
    \caption{(Color Online) Purity critical time $t_c$ as a function
      of $J/\delta$ for different initial states, 
     $n_B=1$ (circles), $n_B=2$ (squares), $n_B=3$ (diamonds), 
     $|{\tt W} \rangle$ (blue triangles,) and different purity measures: 
     $P_1(t)$ (empty) and $P_2(t)$ (full).  
     Inset: $P_2(t)$ vs time in the FGR regime (black
     lines) and in the ergodic regime (red lines) for different
     initial states with $n_B=0,1,2,3$. 
 \vspace*{-2mm}   }
    \label{fig4}
\end{figure}

{\it Initial entangled states.}-- The analysis may be extended to the
dynamics of arbitrarily entangled initial states.  Consider a state in
the central band containing only bipartite entanglement first, for
instance 
\beq |\psi_0 (n_B)\rangle = \overbrace{|01 \dots
01\rangle}^{n-2 n_B} \otimes \left[ \frac{1}{\sqrt 2}(|01\rangle +
|10\rangle) \right]^{\otimes \, n_B}\:,
\label{istate}
\eeq 
where $n_B$ is the number of Bell pairs, $n_B=0$ recovering
the fully separable case.  The local purity evolution is depicted in
Fig.~\ref{fig3}.  Reflecting the fact that $P_1$ is sensitive to all
correlations between spins, the initial value is lower the larger 
$n_B$, ${P}_1(0) = 1-2n_B/n$, using
Eq.~(\ref{local}).  
For $t>0$, $P_1(t)$ decays similarly to the separable case, two 
distinct dynamical regimes emerging for $n_B=0,1,2,3$.  As 
expected, data corresponding to a given regime approximately fall on
the same curve once rescaled by ${P}_1(0)$ (Fig.~\ref{fig3}, inset).
This may be quantitatively understood by studying the evolution of the
bi-local purity defined in Eq. (\ref{bilocal}).  By construction, 
$P_2$ is insensitive to any pairwise correlation 
present in the state (\ref{istate}), effectively mapping the analysis
back to the separable case $n_B=0$.  $P_2(t)$ is
depicted in the inset of Fig.~\ref{fig4}.  Clearly, $P_2(0)=1$
irrespective of $n_B$. The time decay then follows the two regimes
predicted by (\ref{regimes}), again reflecting the underlying
structure of the LDOS (see Fig.~\ref{fig4}).

As a representative example of an initial state containing genuine
multipartite entanglement, we focus on a so-called $|{\tt W}\rangle$
state that is, an equally weighted superposition with equal phases of
the $N_B(1)=n$ states with $z$-magnetization $\sum_j
\sigma_z^{(j)}=n-2$.  Within our disorder model, $|{\tt W}\rangle$ is
automatically protected against the effects of $H_\delta, H_J$ as long
as the coupling between different bands remains small, as assumed so
far. Indeed, the local purity remains almost constant (data not
shown).  Thus, to analyze how the (tripartite) correlations contained
in a $|{\tt W}\rangle$ are affected by static random imperfections we
set $\Delta=0$~\cite{Wnote}. Again, the same qualitative picture
arises, with two distinct dynamical regimes (inset of Fig.~\ref{fig3})
and two different scalings of critical decay times $t_c$
(Fig.~\ref{fig4}).

{\em Conclusions.}-- We established the existence of distinctive
signatures in the dynamics of multipartite entanglement in a quantum
many-body system subject to static disorder.  While a more systematic
analysis is certainly desirable, we expect our main conclusions to
prove valid under more general conditions, including different
dimensionality and/or model Hamiltonians, as well as higher-spin
systems.  Beside reinforcing the usefulness of the GE notion as a
diagnostic framework for complex quantum systems, the deep connections
between GE, LDOS, and fidelity decay emerging from our work are likely
to have broader implications across the fields of quantum information,
condensed-matter physics, and quantum chaos, stimulating crosstalk
between different communities.

We thank W.~G. Brown, R. Fazio, G. Ortiz, L.~F. Santos, and
Y.~S. Weinstein for feedback.  Partial support from the IST-SQUBIT2,
from IBM (Faculty Awards 2005), and from Constance and Walter Burke
through their Special Projects Fund in QIS is acknowledged.

\end{document}